\documentclass[12pt]{article}
\usepackage{amssymb}
\usepackage{epsfig}
\begin{document}
%
\setlength{\baselineskip}{0.65cm}
\setlength{\parskip}{0.35cm}
%
\begin{titlepage}
%
\begin{flushright}
BNL-NT-02/23 \\
RBRC-293 \\
October 2002
\end{flushright}

\vspace*{1.1cm}
\begin{center}
\LARGE

{\bf {Next-to-leading order QCD corrections}}\\

\medskip
{\bf {to high-$\mathbf{p_T}$ pion production in }}\\

\medskip
{\bf { longitudinally polarized $\mathbf{pp}$ collisions}}\\

\vspace*{1.5cm}
\large 
{B.\ J\"{a}ger$^{a}$, A.\ Sch\"{a}fer$^{a}$, M.\ Stratmann$^{a}$, 
and W.\ Vogelsang$^{b,c}$}

\vspace*{1.0cm}
\normalsize
{\em $^a$Institut f{\"u}r Theoretische Physik, Universit{\"a}t Regensburg,\\
D-93040 Regensburg, Germany}\\

\vspace*{0.5cm}
{\em $^b$Physics Department, Brookhaven National Laboratory,\\
Upton, New York 11973, U.S.A.}\\

\vspace*{0.5cm}
{\em $^c$RIKEN-BNL Research Center, Bldg. 510a, Brookhaven 
National Laboratory, \\
Upton, New York 11973 -- 5000, U.S.A.}
\end{center}

\vspace*{1.5cm}
\begin{abstract}
\noindent
We present a calculation for single-inclusive large-$p_T$ pion production in 
longitudinally polarized $pp$ collisions in next-to-leading order QCD. 
We choose an approach where fully analytical expressions for the underlying 
partonic hard-scattering cross sections are obtained.  
We simultaneously rederive the corresponding corrections to unpolarized scattering and 
confirm the results existing in the literature. Our results allow to calculate the 
double-spin asymmetry $A_{LL}^{\pi}$ for this process
at next-to-leading order, which will soon be used at
BNL-RHIC to measure the polarization of gluons in the nucleon.
\end{abstract}
\end{titlepage}
\newpage

\section{Introduction}
%
The measurement of the proton's spin-dependent deep-inelastic 
structure function $g_1^p$ by the EMC~\cite{ref:emc} more than a decade ago
made once again the spin structure of the nucleon an exciting topic,
which deservedly continues to spark large activity by both theorists and
experimentalists.  The original result, that the total quark spin 
contribution to the nucleon spin is only of the order of about $20\%$
has subsequently been confirmed by other experiments and is well-established
today. For various reasons that we will not review here, gluons
may very well play a more important role for the proton
spin than quarks. Consequently, there is now a flurry of experimental
activity aiming at measuring the polarization of gluons in the 
nucleon. In terms of a parton density, the required information
is contained in~\cite{ref:mano}
\begin{equation}
\Delta g \left(x,\mu_F\right) \,=\,
\frac{i}{4\pi\,x\, P^+} \int d\lambda \, {\rm e}^{i\lambda x P^+}\,
\langle P,S| G^{+\nu}(0)\, \tilde{G}^+_{\;\;\nu} 
(\lambda n)|P,S\rangle\Big|_{\mu_F} \;\;,
\end{equation}
written in $A^+=0$ gauge, where $x$ is the gluon's light-cone
momentum fraction of the proton momentum $P^+$, 
and $\mu_F$ the factorization scale
appearing in a hard process to which the gluon contributes.
$G^{\mu\nu}$ is the field strength tensor, and $\tilde{G}^{\mu\nu}$
its dual. In more simple terms, $\Delta g(x,\mu_F)$ describes the
difference in probabilities for finding a gluon with positive
or negative helicity in a proton with positive helicity, at
``resolution'' scale $\mu_F$:
\begin{equation}
\label{eq:pdf}
\Delta g(x,\mu_F) \equiv g_+^+(x,\mu_F) -
                       g_-^+(x,\mu_F)\; ,
\end{equation}
where superscripts (subscripts) denote the proton (gluon) helicity.

Deeply-inelastic scattering (DIS), $lp\to l'X$, is a standard process 
for studying nucleon structure. However, it is not an ideal
process for measuring the gluon content of the nucleon, due to the
fact that the virtual photon in DIS couples directly only to
quarks. Inclusive structure functions therefore depend on the
gluon density only through scale-evolution, and through higher
orders in QCD perturbation theory. This explains why the 
existing polarized-DIS data have told us very little
about $\Delta g$~\cite{ref:grsv,ref:polpdf}.  
One may attempt to get access to the gluon density
by selecting the photon-gluon-fusion process in DIS, which 
contributes to final states such as heavy flavor pairs, or
high-transverse momentum ($p_T$) hadron pairs. Indeed, 
the {\sc Compass} experiment~\cite{ref:compass} at CERN and 
{\sc Hermes}~\cite{ref:hermes} 
at DESY follow
this approach. Unfortunately, the rather low energy in these
fixed-target experiments and the ensuing large systematic 
uncertainties in the theoretical predictions complicate these
efforts significantly. Dedicated experiments at a possibly 
forthcoming future polarized $ep$ collider, like the EIC~\cite{ref:eic}, 
would presumably make these channels more promising, however.

The BNL Relativistic Heavy-Ion Collider RHIC~\cite{ref:rhic} is 
able to run in a mode with polarized protons. 
Very inelastic $pp$ collisions will then open up unequaled
possibilities to measure $\Delta g$.  RHIC has the advantage
of operating at high energies ($\sqrt{S}=200$ and $500$ GeV),
where the theoretical description is under good control. In 
addition, it offers various different channels in which $\Delta
g$ can be studied, such as prompt photon production, jet 
production, creation of heavy flavor pairs, or inclusive-hadron
production. In this way, RHIC is expected to provide the best 
source of information on $\Delta g$ for a long time to come.

The basic concept that underlies most of spin physics at RHIC
is the factorization theorem~\cite{ref:fact}, which states that large
momentum-transfer reactions may be factorized into
long-distance pieces that contain the desired information on the
(spin) structure of the nucleon in terms of its parton
densities such as $\Delta g(x,\mu_F)$, and parts that are
short-distance and describe the hard interactions of the
partons. The two crucial points here are that on the one hand 
the long-distance contributions are universal, i.e., they
are the same in any inelastic reaction under consideration,
and that on the other hand the short-distance
pieces depend only on the large scales related to the 
large momentum transfer in the overall reaction and, therefore,
can be evaluated using QCD perturbation theory. 
The factorized structure forces one to introduce into the calculation
a scale of the order of the hard scale in the reaction -- but not specified 
further by the theory -- that separates the 
short- and long-distance contributions. This scale is the
factorization scale $\mu_F$ mentioned above. 

As an example, let us consider the spin-dependent cross section
for the  reaction $pp\to \pi X$, where the pion is at high
transverse momentum, ensuring large momentum transfer. This is
the reaction we study in the following. The
spin-dependent differential cross section is defined as 
\begin{equation} 
\label{eq1}
d\Delta \sigma \equiv
\frac{1}{2} \left[d\sigma^{++} - d\sigma^{+-}\right] \;\; ,
\end{equation}
where again the superscripts denote the helicities of the 
protons in the scattering. The statement of the 
factorization theorem is then:
\begin{eqnarray}
\label{eq2}
d\Delta \sigma &=&\sum_{a,b,c}\, 
\int dx_a 
\int dx_b 
\int dz_c \,\,
\Delta f_a (x_a,\mu_F) \,\Delta f_b (x_b,\mu_F) 
D_c^{\pi}(z_c,\mu_F') \, \nonumber \\ [2mm]
&&
\times \, d\Delta \hat{\sigma}_{ab}^{c}
(x_aP_A, x_bP_B, P_{\pi}/z_c, \mu_R, \mu_F, \mu_F') \;\; ,
\end{eqnarray}
where the sum is over all  contributing partonic channels $a+b\to
c + X$, with
$d\Delta \hat{\sigma}_{ab}^{c}$ the associated partonic cross
section, defined in complete analogy with Eq.~(\ref{eq1}), the helicities 
now referring to partonic ones: 
\begin{equation} \label{eq3}
d \Delta\hat{\sigma}_{ab}^{c} \equiv
\frac{1}{2} \Bigg[ \left(d\hat{\sigma}_{ab}^{c}\right)^{++} - 
\left(d\hat{\sigma}_{ab}^{c}\right)^{+-}
\Bigg] \;\; .
\end{equation}

A few further comments are in order here. First, Eq.~(\ref{eq2}) is
actually a slight extension of the factorization theorem 
compared to what we stated above: the fact that we are observing
a specific hadron in the reaction requires the introduction of 
additional long-distance functions, the parton-to-pion
fragmentation functions $D_c^{\pi}$. These functions have been 
determined with some accuracy by observing leading pions in
$e^+e^-$ collisions and in DIS.  Even though there is certainly 
room for improvement in our knowledge of the $D_c^{\pi}$, we assume
for this study that the fragmentation functions are sufficiently 
known.

Secondly, we have displayed the full set of required 
scales in Eq.~(\ref{eq2}). Besides the factorization scale $\mu_F$
for the initial-state partons, there is also a factorization
scale $\mu_F'$ for the absorption of long-distance effects into the 
fragmentation functions. In addition, we have a renormalization scale 
$\mu_R$ associated with the running strong coupling constant $\alpha_s$.

As mentioned above, the partonic cross sections may be evaluated
in perturbation theory. Schematically, they can be expanded as
\begin{equation}
d\Delta \hat{\sigma}_{ab}^{c}\,=\,
d\Delta \hat{\sigma}_{ab}^{c,(0)}+
\frac{\alpha_s}{\pi} d\Delta \hat{\sigma}_{ab}^{c,(1)}+\ldots \; .
\end{equation}
$d\Delta \hat{\sigma}_{ab}^{c,(0)}$ is the leading-order (LO)
approximation to the partonic cross section and is, for our
case of pion production, obtained from evaluating all basic $2\to 2$
QCD scattering diagrams. It is therefore of order $\alpha_s^2$.
The lowest order, however, can generally
only serve to give a rough description of the reaction under
study. It merely captures the main features, but does not
usually provide a quantitative understanding. The first-order
(``next-to-leading order'' [NLO]) corrections are generally
indispensable in order to arrive at a firmer theoretical 
prediction for hadronic cross sections.  For instance, the
dependence on the unphysical factorization and renormalization 
scales is expected to be much reduced when going to higher orders in the perturbative
expansion. Only with knowledge of the NLO corrections can one reliably 
extract information on the parton distribution functions from the 
reaction. This is true, in particular, for spin-dependent
cross sections, where both the polarized parton densities and
the polarized partonic cross sections may have zeros in the 
kinematical regions of interest, near which the predictions at
lowest order and the next order will show marked differences.

There has been a lot of effort in recent years~[10-15]
to obtain the NLO corrections for the spin-dependent cross sections 
most relevant for the RHIC spin program. By now, essentially the only 
remaining uncalculated corrections are those for the partonic cross
sections in Eq.~(\ref{eq2}), i.e., inclusive pion production.  
These corrections will be presented in this paper. 
We emphasize that it is very appropriate to provide the NLO 
corrections at this time: it is planned for
the coming RHIC run (early 2003) to attempt a first measurement
of $\Delta g$ through exactly the spin asymmetry 
\begin{equation}
\label{eq:asydef}
A_{LL}^{\pi}=\frac{d\Delta \sigma}{d\sigma}=\frac{ 
d\sigma^{++} - d\sigma^{+-}}{d\sigma^{++} + d\sigma^{+-}}
\end{equation}
for high-$p_T$ 
pion production. The main underlying idea here is that $A_{LL}^{\pi}$ 
is very sensitive to $\Delta g$ through the contributions from
polarized quark-gluon and gluon-gluon scatterings. We note that the 
{\sc Phenix} collaboration has recently presented first, still preliminary, 
results for the 
unpolarized cross section for $pp\to \pi^0 X$ at $\sqrt{S}=200$~GeV, 
which are well described by the NLO QCD calculation~\cite{ref:phenix}, 
providing confidence that the theoretical framework based on perturbative-QCD
hard scattering and summarized by Eq.~(\ref{eq2}) is adequate.

Section~2 gives an outline of the calculation, summarizing the
main ingredients. In Sec.~3 we present some first numerical applications
of our results. 

\section{Calculation of the NLO corrections}
%
\subsection{Outline of the strategy of the calculation}
%
The ``parton-model'' type picture employed in Eq.~(\ref{eq2}) 
implies that the partonic cross sections $d\Delta 
\hat{\sigma}_{ab}^{c}$ are single-inclusive cross
sections for the reactions $a+b\to c+X$, i.e., summed over 
all final states (excluding $c$) possible at the order considered, 
and integrated over the entire phase space of $X$. Writing
out Eq.~(\ref{eq2}) explicitly to NLO, we have
\begin{eqnarray} 
\label{eq1old}
E_{\pi} \frac{d\Delta\sigma}{d^3 p_{\pi}} 
&=& \frac{1}{\pi S} \sum_{a,b,c} \int^1_{z_0} 
\frac{dz_c}{z_c^2} \int_{VW/z_c}^{1-(1-V)/z_c}\!\!\!\frac{dv}{v(1-v)} 
\int^1_{VW/vz_c}\!\!\frac{dw}{w}\Delta f_a(x_a,\mu_F) \Delta f_b(x_b,\mu_F) 
D_c^{\pi}(z_c,\mu_F')  
\nonumber \\[3mm]
&\times&  \,\left[ 
\frac{d\Delta \hat{\sigma}^{c,(0)}_{ab}(v)}{dv} 
\delta (1-w) + \frac{\alpha_s(\mu_R)}{\pi} \, \frac{d\Delta 
\hat{\sigma}^{c,(1)}_{ab}(s,v,w,\mu_R,\mu_F,\mu_F')}{dvdw}
\right] \;\; , 
\end{eqnarray}
where $z_0=1-V+V W$, with hadron-level variables 
\begin{equation}
V\equiv 1+\frac{T}{S} \; \; , \;\;\;\; 
W\equiv \frac{-U}{S+T} \; \; , \;\;\;\; 
S\equiv (P_A+P_B)^2 \; \; , \;\;\;\; T\equiv (P_A-P_{\pi})^2 
\; \; , \;\;\;\; 
U\equiv (P_B-P_{\pi})^2 \;\; ,
\end{equation}
and corresponding partonic ones 
\begin{equation} \label{partvar}
v\equiv 1+\frac{t}{s} \; \; , \;\;\;\; 
w\equiv \frac{-u}{s+t} \; \; , \;\;\;\; 
s\equiv (p_a+p_b)^2 \; \; , \;\;\;\; t\equiv 
(p_a-p_c)^2 \; \; , \;\;\;\; u\equiv (p_b-p_c)^2 \;\; .
\end{equation}
Neglecting all masses, one has the relations
\begin{equation} \label{further}
s=x_a x_b S \;\; , \;\;\;\; t=\frac{x_a}{z_c} T \;\; , \;\;\;\; 
u=\frac{x_b}{z_c} U \;\; , \;\;\;
x_a = \frac{VW}{vwz_c} \;\; , \;\;\;\; x_b = \frac{1-V}{z_c (1-v)} \;\;.
\end{equation}

The LO partonic cross sections $d\Delta \hat{\sigma}^{c,(0)}_{ab}(v)$
are calculated from the $2\to 2$ QCD scattering processes, 
that is, $X$ consists of only one parton, and its phase space is trivial
and leads to the $\delta(1-w)$ factor in Eq.~(\ref{eq1old}).  
We do not need to present the cross sections here, which 
have been known for a long time for both the unpolarized and the polarized 
cases~\cite{ref:bms}. There are actually only four generic reactions, 
$qq'\to qq'$, $qq\to qq$, $q\bar{q}\to gg$, and
$gg\to gg$; all other processes follow from crossing if
one works in terms of helicity amplitudes for each reaction, 
keeping all particles polarized. All tree-level $2\to 2$ helicity
amplitudes are given in~\cite{ref:gw}. The four generic processes
give rise to the ten separate LO channels 
\begin{eqnarray} 
\label{loproc}
qq'&\to& q X \nonumber \\
q\bar{q}'&\to& q X \nonumber \\
q\bar{q}&\to& q' X\nonumber \\
qq&\to& q X\nonumber \\
q\bar{q}&\to& q X\nonumber \\
q\bar{q}&\to& g X\nonumber \\
qg&\to& q X\nonumber \\
qg&\to& g X\nonumber \\
gg&\to& g X\nonumber \\
gg&\to& q X \; ,
\end{eqnarray}
the ``observed'' final-state parton fragmenting into the hadron. 
At NLO, we have ${\cal O}(\alpha_s)$ corrections to the above 
reactions, and also the additional new processes
\begin{eqnarray} \label{nloproc}
qq'&\to& g X \nonumber \\
q\bar{q}'&\to& g X \nonumber \\
qq&\to& g X\nonumber \\
qg&\to& q' X\nonumber \\
qg&\to& \bar{q}' X\nonumber \\
qg&\to& \bar{q} X \; .   
\end{eqnarray}

A single-inclusive-parton cross section 
is, of course, not a priori infrared-finite 
in QCD, but sensitive to long-distance dynamics through the presence
of collinear singularities that arise when the momenta of partons in the 
initial or final states become parallel. Such a situation
can appear for the first time at ${\cal O}(\alpha_s^3)$ (NLO),
where $2\to 3$ scattering diagrams contribute.
From the factorization theorem discussed above it follows that 
long-distance sensitive contributions may be factored into the 
bare parton distribution functions or fragmentation functions. The 
result of this procedure are finite partonic hard-scattering cross 
sections $d\Delta \hat{\sigma}_{ab}^{c}$. At intermediate stages,
however, the calculation will necessarily show singularities
that represent the long-distance sensitivity. In addition,
for those processes that are already present at LO,
real $2\to 3$ and virtual one-loop $2\to 2$ diagrams 
contributing to the calculation will individually have infrared singularities that only cancel
in their sum. Virtual diagrams will also produce ultraviolet
poles that need to be removed by the renormalization of the
strong coupling constant at a scale $\mu_R$. As a result, a regulator has to 
be introduced into the calculation that makes all the 
singularities manifest so that they can be canceled
in the appropriate way. Our choice will be dimensional
regularization, that is, the calculation will be performed
in $d=4-2\varepsilon$ space-time dimensions. Subtractions of
singularities will generally be made in the $\overline{\rm{MS}}$ 
scheme.

Dimensional regularization becomes a somewhat subtle issue 
if polarizations of particles are taken into account. This
is due to the fact that projections on helicities involve the
Dirac matrix $\gamma_5$ for quarks and the Levi-Civita
tensor $\epsilon^{\mu\nu\rho\sigma}$ for gluons. These
two objects are genuinely four-dimensional and hence do not have
a natural extension to $4-2\varepsilon$ dimensions. In fact, some
care has to be taken to avoid algebraic inconsistencies in the
calculation when using $\gamma_5$ and $\epsilon^{\mu\nu\rho\sigma}$.
At the level of the algebra the treatment of $\gamma_5$ 
and $\epsilon^{\mu\nu\rho\sigma}$ of course only affects terms that
are of ${\cal O}(\varepsilon)$. However, poles proportional to $1/\varepsilon$ and 
$1/\varepsilon^2$ present in the calculation may combine with
this to eventually result in non-vanishing contributions. 
A widely-used scheme for dealing with $\gamma_5$ 
and $\epsilon^{\mu\nu\rho\sigma}$ in a fully 
consistent way is the one developed in~\cite{ref:hvbm}, the HVBM scheme. 
This is the scheme we have used for our calculation.
It is mainly characterized by splitting the $d$-dimensional
metric tensor into a four-dimensional and a $(d-4)$-dimensional
one. The Levi-Civita tensor is then defined by having components
within the four-dimensional subspace only, and $\gamma_5$ 
anti-commutes with the other Dirac matrices in the four-dimensional
subspace, but commutes with them in the $(d-4)$-dimensional one.
The HVBM scheme leads to a higher complexity of the algebra and of
phase space integrals. However, one may make use of computer algebra
programs such as Tracer~\cite{ref:tracer} that allow to handle the 
split-up of space-time, and the treatment of $(d-4)$-dimensional
components in phase space integrals has become rather standard by now. 
We emphasize that for our present case the treatment of 
$\gamma_5$ and $\epsilon^{\mu\nu\rho\sigma}$ has no bearing
on the ultraviolet (renormalization) sector of the calculation,
since we have no chiral vertices in the calculation. For
instance, we may perform all renormalizations at the level
of vertex and self-energy diagrams, without reference to the
polarizations of the external particles.

As remarked above, we need to sum over all possible final states
in each channel $ab\to c X$, in compliance with the requirement 
of single-inclusiveness of the cross section. For instance, in case of 
$qg\to qX$ one needs, besides the virtual corrections to $qg\to qg$,
three different $2\to 3$ reactions: $qg\to q(gg)$, $qg\to q(q\bar{q})$,
$qg\to q(q'\bar{q}')$ (where brackets indicate the unobserved parton
pair). Only all three processes combined will allow to arrive at
a finite answer in the end. The summation over $X$ is therefore always 
implicitly understood in the following. 

In addition, the two unobserved partons in the $2\to 3$ contributions 
need to be integrated over their entire phase space. 
The integration may be performed in basically two different ways. 
The first one relies on Monte-Carlo integration techniques. As
was shown in~\cite{ref:fks,ref:ks}, the regions where the squared $2\to 3$  
matrix elements become singular can be straightforwardly identified 
and separated. 
These regions will yield all the poles in $1/\varepsilon$ after
integration, which eventually must cancel as described above. It then
becomes possible to organize the calculation in such a way that
the singularities are extracted and canceled by hand, while the 
remainder may be integrated numerically over phase space. This
approach has the advantage of being very flexible; it may
be used for any infrared-safe observable, with any experimental 
cut~\cite{ref:fks}. On the other hand, the numerical integration involved 
turns out to be rather delicate and time-consuming. In case of polarized 
collisions, the method was employed for the reactions 
$pp\to {\rm jet}X$~\cite{ref:ddfsv} 
and $pp\to \gamma X$~\cite{ref:fv} at NLO.

The method we will employ is to perform the phase space integration 
of the $2\to 3$ contributions {\em analytically}. This has several 
advantages. In first place, the final answer is much more amenable
to a numerical evaluation, giving much more stable results 
in a much shorter time. This may become important at a later
stage, when experimental data will have been obtained and one is aiming
to extract $\Delta g$ from them within a ``global analysis''~\cite{ref:sv}. 
In addition, the ``analytic method'' has also been employed in the
unpolarized case~\cite{ref:aversa}. Since the calculation
of the unpolarized and the polarized NLO terms largely proceeds
along similar lines, we can compute both simultaneously. 
Our results for the unpolarized case may then be compared at an
{\em analytical level} to those available in the numerical code of~\cite{ref:aversa}, which 
provides an extremely powerful check on the correctness of all our
calculations. 

We will now separately address the virtual $2\to 2$ and 
real-emission $2\to 3$ NLO contributions. Then we will
discuss their sum and the cancellation of singularities.

\subsection{Virtual contributions}
\begin{figure}[t]
\begin{center}
\epsfig{figure=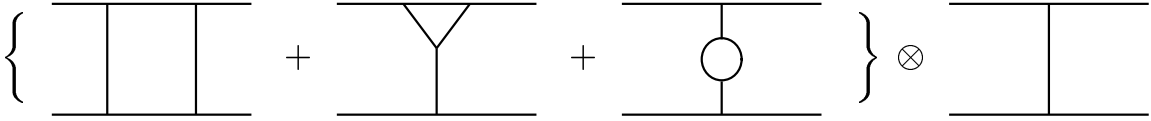,width=0.95\textwidth}
\end{center}
\vspace*{-1.0cm}
\caption{\sf Interference of generic virtual (box, vertex,
self-energy) contributions with Born diagrams. \label{fig:fig1}}
\vspace*{-0.5cm}
\end{figure}

%
At ${\cal O}(\alpha_s^3)$, virtual corrections only contribute
through their interference with the Born diagrams, as sketched
in Fig.~\ref{fig:fig1}. We have calculated the virtual contributions
with two different methods.
%

Firstly, we have performed a direct calculation. Here 
we could make use of known $\overline{\rm{MS}}$-renormalized 
one-loop vertex and self-energy structures as given in~\cite{ref:nps}, 
which may be readily inserted into the Born diagrams. One then 
additionally needs to calculate the box diagrams which are ultraviolet
finite and hence not subject to the renormalization procedure. We have
simultaneously computed the virtual corrections for the
unpolarized case and found complete agreement with the 
results published in~\cite{ref:es}. 

The second approach makes use of the fact that in Ref.~\cite{ref:kst}
the helicity amplitudes for all one-loop $2\to 2$ QCD scattering 
diagrams were presented. It is clear that these contain the
information we need for our calculation. The only subtlety is
that the helicity method employed in~\cite{ref:kst} will not immediately
yield the answer for the HVBM prescription we are looking for. However, 
as was also pointed out in~\cite{ref:ks,ref:kst}, the translation between the 
results for the two schemes is fairly straightforward. In fact, by 
inspecting the singularity structure of the diagrams, one can derive a 
universal form for the virtual contribution ${\cal V}$ 
that schematically reads:
\begin{equation}
\label{Kunsztform}
{\cal V}(s,t,u) = {\cal B}(s,t,u)
\left\{-\frac{1}{\varepsilon^2}\sum_n C_n
-\frac{1}{\varepsilon}\sum_n \gamma_n\right\} 
+\frac{1}{\varepsilon}\sum_{m<n}
\log\left(\frac{2 p_n\cdot p_m}{s}\right)
\tilde{{\cal B}}_{mn}(s,t,u)
+\tilde{{\cal V}}(s,t,u),
\end{equation}
where $n,m$ are summed over all external legs, the $p_i$ are the external parton momenta, 
and ${\cal B}$ denotes the Born cross section for the reaction under consideration. The
$\tilde{{\cal B}}_{mn}$ are the so-called ``color-linked''
Born cross sections, to be calculated 
according to rules given in~\cite{ref:ks}. The $C_i$ and $\gamma_i$ 
are coefficients depending only on the type of external leg, with 
$C_q= C_F=4/3$, $C_g = C_A=3$, $\gamma_q = 3C_F/2$, 
$\gamma_g =\beta_0/2=11/2-n_f/3$, $n_f$ being the number of flavors.
Finally, $\tilde{{\cal V}}$ is the finite remainder. The only
difference between the result for the virtual correction 
in the helicity amplitude method and the conventional HVBM scheme 
resides in the ${\cal B}$ and $\tilde{{\cal B}}_{mn}$ terms. 
For the helicity method, these are four-dimensional quantities,
whereas in conventional dimensional regularization they 
are calculated in $d$ dimensions in the HVBM scheme. This property
allows a direct determination of the full virtual correction in the 
HVBM scheme, since ${\cal V}$ has been calculated with helicity 
amplitude methods in~\cite{ref:ks}. This strategy for determining the 
virtual corrections was also adopted in Ref.~\cite{ref:ddfsv}.
 
We found complete agreement between the results obtained for
our two approaches for obtaining the virtual corrections.
%
\begin{figure}[t]
\begin{center}
\epsfig{figure=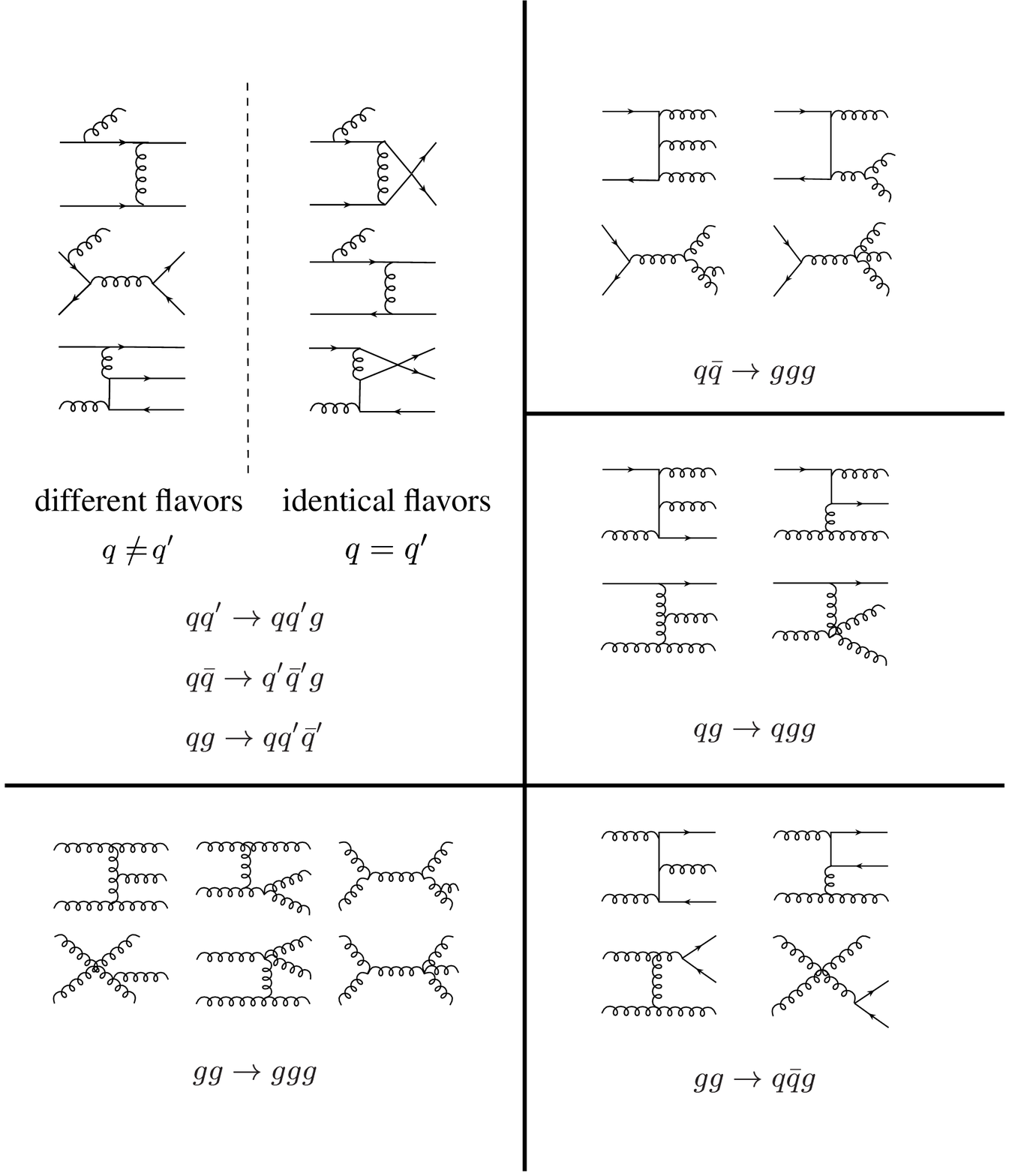,width=0.75\textwidth}
\end{center}
\caption{\sf Representative $2\to 3$ Feynman diagrams contributing to $a b\to cX$
to ${\cal{O}}(\alpha_s^3)$. \label{fig2}}
\end{figure}
%

\subsection{$\mathbf{2\to 3}$ real contributions}
%
Figure~\ref{fig2} shows some representative $2\to 3$ 
Feynman diagrams contributing to $a b\to cX$
to ${\cal{O}}(\alpha_s^3)$. The squared spin-dependent 
matrix elements in $d$ dimensions, using the 
HVBM prescription for $\gamma_5$ and the Levi-Civita tensor, are
too lengthy to be reported here. Again, we have simultaneously
calculated the squared matrix elements for the unpolarized
case, and we recover the known~\cite{ref:es} results in $d$ dimensions.
The polarized matrix elements can be checked in $d=4$ dimensions
against the expressions in~\cite{ref:gw}, and again we find agreement. 

In $d=4-2\varepsilon$ dimensions, as a consequence of using the HVBM scheme
with its distinction between four- and $(d-4)$-dimensional subspaces,
the squared matrix elements contain scalar products of vectors 
separately in these subspaces. For instance, while an outgoing
unobserved parton with momentum $k$ is massless, $k^2=0$, 
we may encounter its $(d-4)$-dimensional invariant mass, 
denoted as $\hat{k}^2$, in the calculation, which is constructed from the
$(d-4)$-dimensional components of $k$. Such terms need
to be carefully taken into account in the phase space integrations.

The most economical way of organizing the phase space integration
is to work in the rest frame of the two unobserved final-state partons
whose momenta $k_2$ and $k_3$ can then be parameterized as
\begin{eqnarray}
\label{eq:k2k3}
k_2 &=& (k_0,k_0 \sin \theta_1 \cos \theta_2, k_y, k_0 \cos \theta_1, \hat{k})\;\;, \nonumber \\
k_3 &=& (k_0,-k_0 \sin \theta_1 \cos \theta_2, -k_y, -k_0 \cos \theta_1, -\hat{k})\;\;,
\end{eqnarray}
and to define the momenta of the other three particles 
to lie in the $x-z$ plane in the four-dimensional space. In this case
the above $\hat{k}^2$ is the only invariant arising from the
$(d-4)$-dimensional subspace. In Eq.~(\ref{eq:k2k3}) 
$k_0=\sqrt{s_{23}}/2$ with $s_{23}=(k_2+k_3)^2=s\,v\, (1-w)$, and $k_y$ denotes the
unspecified $y$ component of $k_2$ and $k_3$ which can be trivially integrated over since the matrix element
does not depend on it. One then has the three-body phase space~\cite{ref:gv}
\begin{eqnarray}
\lefteqn{
\Phi_3 = \frac{s}{(4\pi)^4 \Gamma (1-2\varepsilon)} \left(
\frac{4\pi}{s} \right)^{2\varepsilon} \int_0^1 dv v^{1-2 \varepsilon}
(1-v)^{-\varepsilon} \int_0^1 dw \left( w (1-w) \right)^{-\varepsilon} } 
\nonumber \\[3mm]
&\times& \int_0^{\pi} d\theta_1 \sin^{1-2\varepsilon} \theta_1
\int_0^{\pi} d\theta_2 \sin^{-2\varepsilon} \theta_2 \,\frac{1}{B(\frac{1}{2},
-\varepsilon)}\int_0^1 \frac{dx}{\sqrt{1-x}} x^{-(1+\varepsilon)} \:\: ,
\end{eqnarray}
where $x$ is $\hat{k}^2$ normalized to its upper limit,
$ x\equiv 4 \hat{k}^2/s_{23} \sin^2\theta_1 \sin^2\theta_2$.

The integrations we do analytically are over $x$ (for those
terms in the squared matrix elements that have dependence
on $\hat{k}^2$) and the angles $\theta_1$ and $\theta_2$. 
$v$ and $w$, defined in Eq.~(\ref{partvar}),
become integration variables in the convolution with the
parton densities, according to Eqs.~(\ref{eq1old}) 
and~(\ref{further}). Extensive 
partial fractioning of the squared matrix elements always leads to 
the master integral for the angular integrations
\begin{eqnarray} \label{ps}
\lefteqn{   \!\!\! \!\!\! \!\!\!\!\!\!
\int_0^{\pi}d\theta_1 \sin^{1-2 \varepsilon} \theta_1
\int_0^{\pi}d\theta_2 \sin^{-2 \varepsilon} \theta_2
\frac{1}{(1-\cos \theta_1)^j (1-\cos \theta_1 \cos \chi
-\sin \theta_1 \cos \theta_2 \sin \chi)^l } } \nonumber \\[3mm]
&=& 2\pi \frac{\Gamma (1-2\varepsilon)}{\Gamma (1-\varepsilon )}
2^{-j-l}\frac{B(1-\varepsilon-j,1-\varepsilon-l)}{\Gamma (1-\varepsilon )}
\: {}_2 F_1 (j,l,1-\varepsilon;\cos^2 \frac{\chi}{2}) \;\; ,
\end{eqnarray}
where the last line is the result given in Ref.~\cite{ref:neerv}.
$B(x,y)$ is the Euler Beta function and $_2 F_1(a,b,c;z)$ denotes the
hypergeometric function.

The final step in the evaluation of the $2\to 3$ contributions
is to extract the poles arising when the invariant mass
of the unobserved partons becomes small: $s_{23}\to 0$. 
According to Eq.~(\ref{partvar}) or the definition of $s_{23}$ below
Eq.~(\ref{eq:k2k3}),
this is the case for $w\to 1$. The fact that the LO contribution
is proportional to $\delta (1-w)$ indicates that the dependence
on $w$ is in the sense of a mathematical distribution. At NLO,
the integrated matrix elements have terms proportional to $1/(1-w)$ which,
after inclusion of the factor $(1-w)^{-\varepsilon}$ from the 
phase space integral~(\ref{ps}), can be expanded as
\begin{equation}
\left( 1-w \right)^{-1-\varepsilon} = -\frac{1}{\varepsilon} \delta (1-w)+
\frac{1}{(1-w)_+}-\varepsilon \left( \frac{\ln (1-w)}{1-w} \right)_+
+{\cal O}(\varepsilon^2) \;\; ,
\end{equation}
making the singularities at $\varepsilon=0$ manifest. Here
the ``$+$''-distributions are defined in the usual way, 
\begin{equation}
\int_0^1 f(w)\left[g(w)\right]_+\,dw =\int_0^1 \left[f(w)-f(1) 
\right] g(w)\, dw \;\; .
\end{equation}

\subsection{Cancellation of singularities}
%
As mentioned earlier, genuine infrared singularities cancel in the
sum of virtual and real contributions, among them all poles
proportional to $1/\varepsilon^2$ which arise when soft and collinear 
singularities coalesce. The sum of virtual and real pieces
is still singular for $\varepsilon \to 0$ as a result of collinear 
divergencies. These remaining poles need to be factored 
into the bare parton distribution functions and fragmentation
functions, depending on whether their origin was in the
initial or final state. 

\begin{figure}[t]
\begin{center}
\epsfig{figure=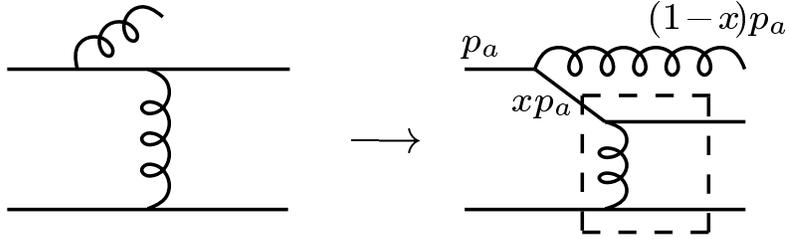,width=0.65\textwidth}
\end{center}
\vspace*{-1cm}
\caption{\sf Representative collinear contribution to the subprocess $qq\to qqg$
(see text). \label{fig:fig3}}
\end{figure}

Figure~\ref{fig:fig3} sketches a typical collinear situation in a
$2\to 3$ process. The contribution displayed will require
a subtraction of the form $\propto \frac{1}{\varepsilon} 
\Delta P_{qq} \times \Delta \hat{\sigma}_{qq\to qq}$,
where $\Delta P_{qq}$ is the spin-dependent LO $q\to q$ Altarelli-Parisi splitting
function~\cite{ref:ap} and $\Delta \hat{\sigma}_{qq\to qq}$ 
represents the subsequent polarized LO scattering $qq\to qq$, 
evaluated in $d$ dimensions. More precisely, 
the structure of this particular collinear subtraction is
\begin{equation}
-\frac{\alpha_s}{\pi} \, \int_0^1 dx\;\Delta H_{qq}(x,\mu_F)\,
\frac{d\Delta \hat{\sigma}_{qq\to qq}(x s,x t,u,
\varepsilon)}{dv}\,\delta \left( x\;(s+t )+u \right)  \;\; ,
\end{equation}
where 
\begin{equation} \label{subfac}
\Delta H_{qq} (z,\mu_F) \equiv \left(-\frac{1}{\varepsilon}
+\gamma_E-\ln 4\pi \right) \Delta
P_{qq} (z) \left( \frac{s}{\mu_F^2} \right)^{\varepsilon} +
\Delta f_{qq} (z)  \;\; .
\end{equation}
Here the Euler constant $\gamma_E$ and $\ln 4\pi$ are the
terms that are commonplace to subtract in order to work
in the $\overline{\rm{MS}}$ scheme. $\Delta f_{qq} (z)$ is an
additional finite piece in the subtraction that represents the freedom 
in choosing a factorization prescription and will be discussed below.
We see in~(\ref{subfac}) how the factorization
scale $\mu_F$ emerges in the subtraction. In general, a process at NLO
will require several collinear subtractions, in both the 
initial and the final states. Depending on which types of partons 
are collinear, the other splitting functions $\Delta P_{qg}$, 
$\Delta P_{gq}$, $\Delta P_{gg}$, as well as other $2\to 2$ cross
sections, will contribute. In the final-state collinear case, a
singularity occurs when the observed parton and an unobserved
one become collinear. The subtraction needed here can be easily
written down in a form analogous to~(\ref{subfac}); it will
involve the final-state factorization scale $\mu_F'$. Note that,
since we are not considering polarization in the final state,
only spin-independent splitting functions appear in the final-state
factorization subtraction.

Taking the $\overline{\rm{MS}}$ scheme literally, one would not have
any additional finite pieces in the subtraction, beyond those
involving $\gamma_E$ and $\ln 4\pi$. That is, one would define
$(\Delta) f_{ij} (z)=0$ in the functions $(\Delta) H_{ij}$ involved
in the various subtractions in the polarized and unpolarized cases.
However, there is a well-known~\cite{ref:gv,ref:mvn,ref:wv} subtlety arising
in the $q\to q$ splitting in the polarized case that is related to
the use of the HVBM scheme. It is a property of the HVBM-scheme definition
of $\gamma_5$ that it leads to helicity non-conservation at the $qqg$ 
vertex in $d$ dimensions. This can be seen from the non-vanishing 
difference of unpolarized and polarized $d$-dimensional LO quark-to-quark 
splitting functions:
\begin{equation} \label{pqqdim}
\Delta P_{qq}^{4-2 \varepsilon} (x)-P_{qq}^{4-2 \varepsilon} (x) =
4 C_F\, \varepsilon\, (1-x) \;\; .
\end{equation}
%
%
A disagreeable consequence of this is a non-zero first moment 
($x$-integral) of the flavor non-singlet NLO anomalous dimension 
for the evolution of spin-dependent quark densities, in conflict with the 
conservation of flavor non-singlet axial currents~\cite{ref:mvn,ref:wv,ref:svw}. 
It is therefore advisable -- albeit not mandatory in a purely mathematical 
sense -- to slightly deviate from the $\overline{\rm{MS}}$ scheme in the 
polarized case by choosing~\cite{ref:mvn,ref:wv}
\begin{equation} \label{fqq}
\Delta f_{qq} (z) = -4 C_F (1-z) \;\; .
\end{equation}
It is important to point out that the choice of the function
$\Delta f_{qq} (z)$ corresponds to the freedom in defining a
factorization scheme. Of course, a physical quantity like the
pion production cross section must not depend on the convention
regarding which finite terms are subtracted from the partonic
cross sections along with the collinear poles. Indeed, the
parton distributions functions are scheme dependent
as well, so that at any given order in $\alpha_s$ the scheme 
dependence cancels in the physical observable. The factorization 
scheme defined by the choice~(\ref{fqq}) has also been used in the 
available sets of spin-dependent NLO $\overline{\mathrm{MS}}$ parton densities, so our definition 
is consistent with these densities. Since the HVBM ``$\gamma_5$-effect'' 
mentioned above is to be regarded as an artifact of the prescription 
and may be removed in a straightforward way by exploiting the conservation 
of non-singlet axial currents, results of polarized NLO calculations are 
usually regarded as being ``genuinely'' in the conventional 
$\overline{\rm{MS}}$ scheme only {\em with} the choice~(\ref{fqq}).
All other possible $\Delta f_{ij}$ are, however, set to zero, as in the 
usual $\overline{\rm{MS}}$ scheme. Needless to say that in the unpolarized 
case one has $f_{qq}=f_{qg}=f_{gq}=f_{gg}=0$
in $\overline{\rm{MS}}$.  

\subsection{Final results}
%
Once we have performed the factorization of collinear singularities,
we arrive at the final result for the NLO partonic hard-scattering
cross sections. We first of all note that, as mentioned earlier,
we have calculated in parallel the NLO corrections for the
unpolarized case. We have compared them term-by-term with the
known analytical results in the code of~\cite{ref:aversa} and found
complete agreement. 

Our results for the spin-dependent NLO corrections may for
each of the 16 subprocesses be cast into the following form:
\begin{eqnarray}
&&\hspace*{-1cm}s\, \frac{d \Delta \hat{\sigma}_{ab}^{c,(1)}
(s,v,w,\mu_R,\mu_F,\mu_F')}{dvdw} = \left(
\frac{\alpha_s (\mu_R)}{\pi} \right)^2
  \left[ \left(  A_0  \delta (1-w) +   B_0
\frac{1}{(1-w)_+} +  C_0 \right) \ln \frac{\mu_F^2}{s} \right. 
\nonumber \\[3mm]
&&+\left( A_1  \delta (1-w) +   B_1
\frac{1}{(1-w)_+} +  C_1 \right) \ln \frac{\mu_F'^2}{s} +
A_2 \delta (1-w) \ln \frac{\mu_R^2}{s} 
\nonumber \\[3mm]
&& + A \delta (1-w)  +B \frac{1}{(1-w)_+} +  C + D\left(
\frac{\ln (1-w)}{1-w} \right)_+ + E \ln w + F \ln v \nonumber  \\[3mm]
&&+ G \ln (1-v) + H \ln(1-w)  + I  \ln (1-vw)+
J  \ln (1-v+vw)  \nonumber \\[3mm]
&&\left.+K \frac{\ln w}{1-w} + L \frac{\ln \frac{1-v}{1-vw}}{1-w}
 +M \frac{\ln (1-v+vw)}{1-w}  \right] \:\:\: ,
\label{final}
\end{eqnarray}
where all coefficients are functions of $v$ and $w$, except those
multiplying the distributions $\delta(1-w)$, $1/(1-w)_+$, 
$\left[ \ln(1-w)/(1-w)\right]_+$ which may be written as 
functions just of $v$. Terms with distributions are present only
for those subprocesses that already contribute at the Born level,
see Eq.~(\ref{loproc}).

We finally make a few observations about our results for the 
polarized case. Consider, for example, the subprocess
$q\bar{q}\to q'X$ in Eq.~(\ref{loproc}). All Feynman diagrams
contributing to this cross section at NLO, virtual as well as real, 
are annihilation diagrams, meaning that the initial quark and antiquark
legs are part of the same fermion line. Independently of the number
of gluons attaching to the fermion line, helicity conservation in QCD
demands that the annihilation can only occur if the quark and
antiquark have opposite helicities. Keeping in mind the 
definition~(\ref{eq3}) for the polarized cross section, we are led
to the expectation that
\begin{equation} \label{helcon}
d \Delta \hat{\sigma}_{q\bar{q}}^{q',(1)}\equiv - 
d\hat{\sigma}_{q\bar{q}}^{q',(1)}
\end{equation}
should be fulfilled for this process. The only way in which this 
relation could be broken is if the regularization we adopt in
the NLO calculation does not respect helicity conservation. As
we discussed earlier, the HVBM prescription for $\gamma_5$ indeed has
this deficiency. However, as known from~\cite{ref:mvn,ref:wv}, the additional 
finite term $\Delta f_{qq}$~(\ref{fqq}) in the factorization
subtraction~(\ref{subfac}) is precisely designed to cure this shortcoming 
of the HVBM prescription and to restore helicity conservation. This
is probably the most tangible reason why the choice Eq.~(\ref{fqq}) is 
required from a physical point of view. The implication of this
is that our final results for $q\bar{q}\to q'X$ should indeed 
satisfy~(\ref{helcon}), which we have verified. We can actually
go one step further: the channels $q\bar{q}\to gX$ and $q\bar{q}\to qX$ 
have contributions from annihilation diagrams as well, but also ones from 
non-annihilation diagrams, for which the $q$ and $\bar{q}$ scatter
via $t$-channel gluon exchange. Helicity conservation makes no immediate 
statement about the non-annihilation diagrams. However, the 
channels $q\bar{q}'\to gX$ and $q\bar{q}'\to qX$ are described by the
non-annihilation diagrams {\em alone}. If we subtract the corresponding 
cross sections from the ones for $q\bar{q}\to gX$ and $q\bar{q}\to qX$,
respectively, we can use helicity conservation again for the remainder. 
Explicitly, we expect:
\begin{equation} \label{helcon1}
\left[ d \Delta \hat{\sigma}_{q\bar{q}}^{q,(1)} - 
d \Delta \hat{\sigma}_{q\bar{q}'}^{q,(1)} \right]\equiv -
\left[ d  \hat{\sigma}_{q\bar{q}}^{q,(1)} - 
d  \hat{\sigma}_{q\bar{q}'}^{q,(1)} \right] \; ,
\end{equation}
and similarly for an observed gluon. Again we have verified that
our final results obey this relation, which we consider a very
powerful check on the correctness of our results.

\section{Numerical results}
%
In this Section, we present a first numerical application of our
analytical results. Instead of presenting a full-fledged phenomenological study
of single-inclusive hadron production in polarized $pp$ collisions, 
which we leave for a future study, we only report the main features of
the NLO corrections and describe their impact on the
cross sections and the spin asymmetry $A_{LL}^{\pi^0}$. 
Predictions for $A_{LL}^{\pi^0}$ are in immediate demand for an extraction 
of $\Delta g$ at RHIC in the very near future.

For our calculations we assume the same kinematic coverage as in the recent 
{\sc Phenix} measurement of the unpolarized
cross section at $\sqrt{S}=200\,\mathrm{GeV}$~\cite{ref:phenix}, that is, we consider pion
transverse momenta in the range $2\leq p_T \leq 13$ GeV
and pseudorapidities $|\eta|\leq 0.38$. We also take into account
that the pion measurement is at present possible only over half the azimuthal 
angle.

We will evaluate cross sections and spin asymmetries at both LO and
NLO, in order to study the size and importance of the corrections we have calculated. 
We will always perform the NLO (LO) calculations using NLO (LO)
parton distribution functions, fragmentation functions, and 
the two-loop (one-loop) expression for $\alpha_s$. 
To calculate the NLO (LO) unpolarized pion cross section needed for the 
denominator of the spin asymmetry $A_{LL}^{\pi^0}$ in Eq.~(\ref{eq:asydef}), we use the 
CTEQ5M (CTEQ5L)~\cite{ref:cteq5m} parton distribution functions. 
In all our calculations we use the pion fragmentation functions
of Ref.~\cite{ref:kkp}, which provides both a LO and a NLO set.
For the polarized cross section, we will mainly use the (NLO/LO) ``standard'' 
sets of the  spin-dependent GRSV~\cite{ref:grsv} parton distributions
(``GRSV-std'').
Since we also want to investigate the sensitivity of $A_{LL}^{\pi^0}$
to the polarized gluon density $\Delta g$, we use another set
of GRSV distributions, for which the gluon is assumed to be 
particularly large (``GRSV-max''). We note that the value of
the strong coupling $\alpha_s$ to be used in conjunction with the 
unpolarized parton distributions differs from that employed in the fits
for the polarized sets and the fragmentation functions.  Our
convention will be to calculate the cross sections always with 
the strong coupling constant accompanying the parton distributions
used. 

\begin{figure}[t]
\vspace*{-0.5cm}
\begin{center}
\epsfig{figure=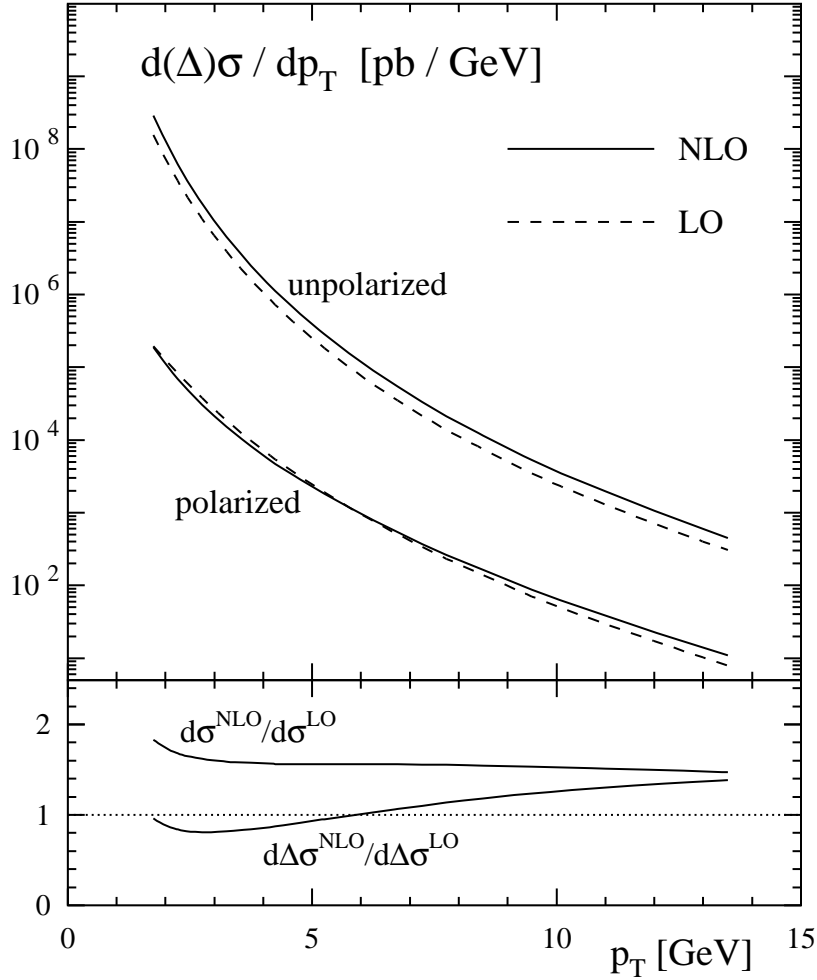,width=0.75\textwidth}
\end{center}
\vspace*{-1.0cm}
\caption{\sf Unpolarized and polarized $\pi^0$ production cross sections in 
NLO (solid) and LO (dashed) at $\sqrt{S}=200$ GeV. The lower panel 
shows the ratios of the NLO and LO results in each case.  \label{fig4}}
\end{figure}
Figure~\ref{fig4} shows our results for the unpolarized and polarized
cross sections at NLO and LO, where we have chosen the scales
$\mu_R=\mu_F=\mu_F'=p_T$. The lower part of the figure displays
the ``$K$-factor''
\begin{equation}
K=\frac{d(\Delta)\sigma^{\rm NLO}}{d(\Delta)\sigma^{\rm LO}} \;\; .
\end{equation}
One can see that in the unpolarized case the corrections are roughly 
constant and about $50\%$ over the $p_T$ region considered. 
In the polarized case, we find
generally smaller corrections which become of similar size as those
for the unpolarized case only at the high-$p_T$ end. The cross section for
$p_T$ values smaller than about 2 GeV is outside the domain of 
perturbative calculations as indicated by rapidly increasing
NLO corrections and, therefore, is not considered here.

%
\begin{figure}[t]
\vspace*{-0.5cm}
\begin{center}
\epsfig{figure=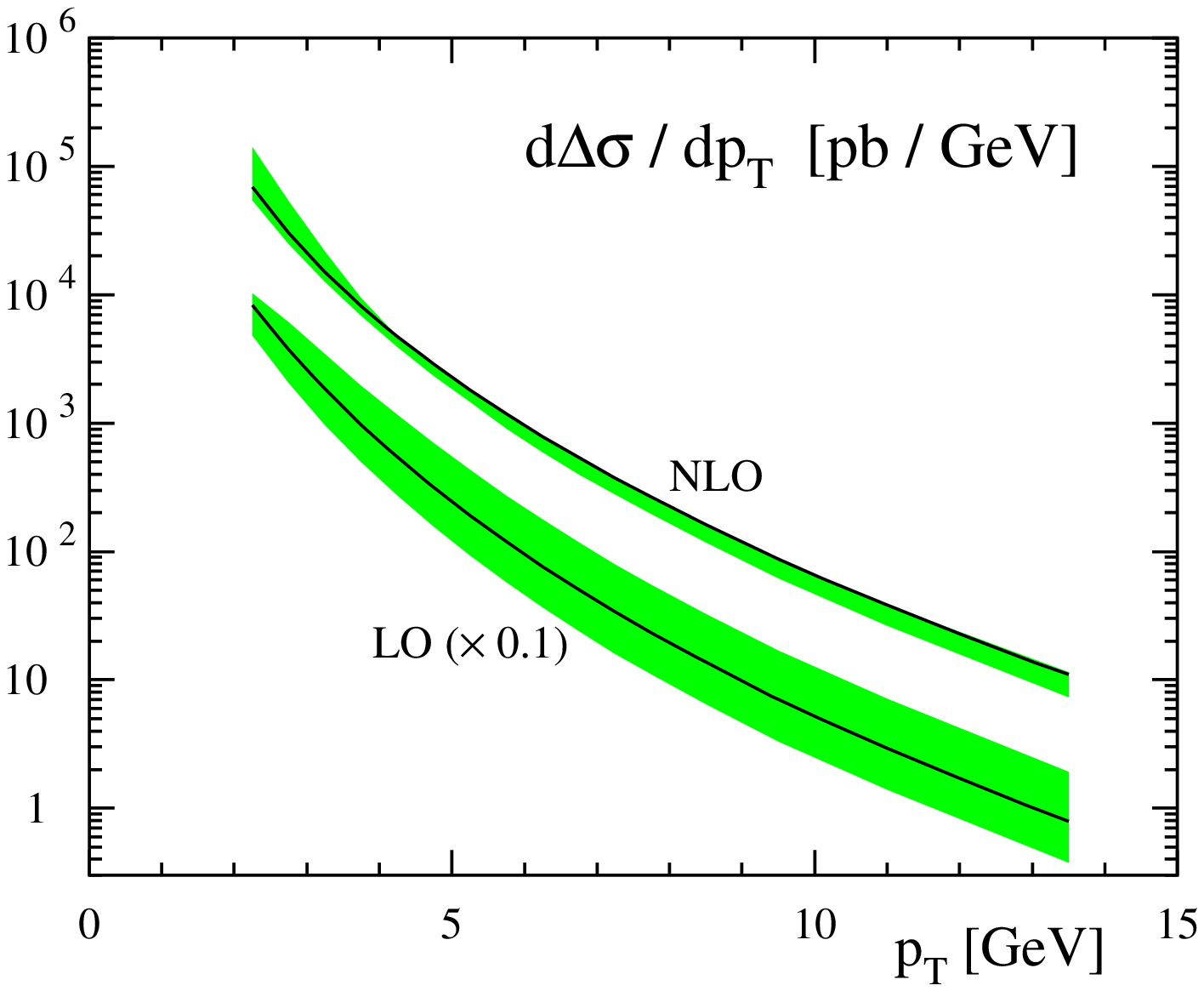,width=0.80\textwidth}
\end{center}
\vspace*{-1.0cm}
\caption{\sf Scale dependence of the polarized cross section
for $\pi^0$ production at LO and NLO in the range 
$p_T/2 \le \mu_R=\mu_F=\mu_F' \le 2 p_T$. We have rescaled the LO results by $0.1$ to 
separate them better from the NLO ones. In each case the solid line corresponds
to the choice where all scales are set to $p_T$.\label{fig5}}
\end{figure}
As we have mentioned in the Introduction, one reason why NLO corrections are 
generally important is that they should considerably reduce the dependence of the cross
sections on the unphysical factorization and renormalization scales. In this
sense, the $K$-factor is actually a quantity of limited significance 
since it is likely to be rather scale dependent through the presence
of the LO cross section in its denominator. The improvement in
scale dependence when going from LO to NLO is, therefore, a 
better measure of the impact of the NLO corrections, and, perhaps,
provides also a rough estimate of the relevance of even higher order QCD corrections. 
Figure~\ref{fig5} shows the scale dependence of the spin-dependent
cross section at LO and NLO. 
In each case the shaded bands indicate the uncertainties from varying the
unphysical scales in the range $p_T/2 \le \mu_R=\mu_F=\mu_F' \le 2 p_T$. The solid
lines are for the choice where all scales are set to $p_T$.
One can see that the scale dependence indeed becomes much smaller at NLO.
 
\begin{figure}[t]
\vspace*{-0.5cm}
\begin{center}
\epsfig{figure=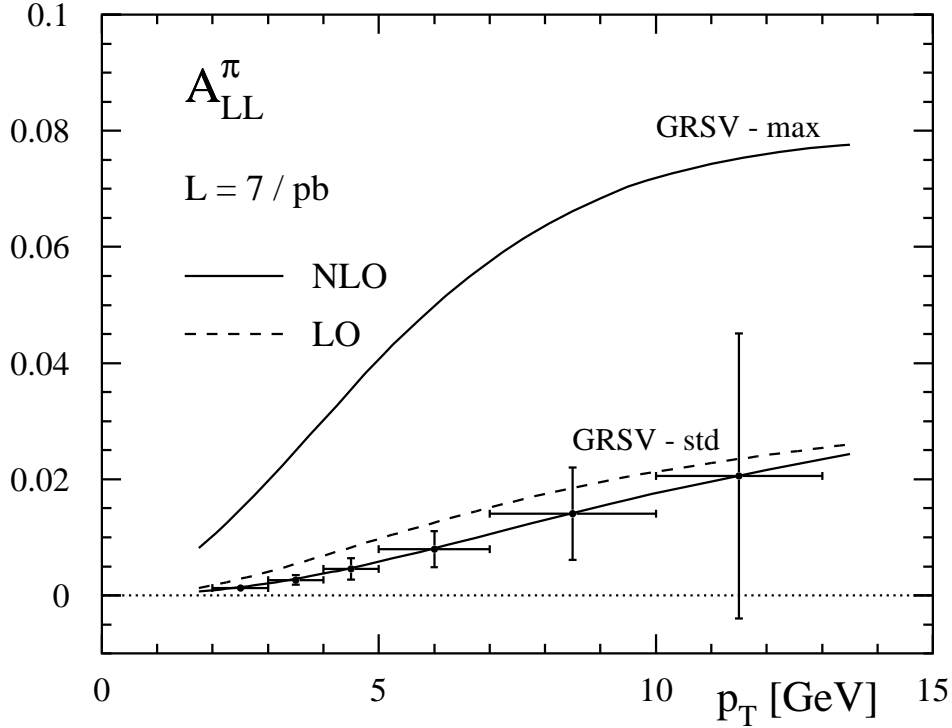,width=0.85\textwidth}
\end{center}
\vspace*{-1cm}
\caption{\sf Spin asymmetry for $\pi^0$ production, using the ``standard''
set of GRSV~\cite{ref:grsv} and the one with ``maximal'' gluon polarization. The
dashed line shows the asymmetry at LO for the GRSV ``standard''
set. The ``error bars'' indicate the expected statistical accuracy targeted
for the upcoming run of RHIC (see text). \label{fig6}}
\end{figure}

Finally, we consider the spin asymmetry which is the main 
quantity of interest here. Figure~\ref{fig6} shows $A_{LL}^{\pi^0}$,
calculated at NLO (solid lines) for the ``standard'' set of GRSV 
parton distributions, and for the one with ``maximal'' gluon 
polarization~\cite{ref:grsv}. We have again chosen all scales to be $p_T$. 
For comparison, we also show the LO result for the GRSV ``standard''
set (dashed line). As expected from the larger $K$ factor for
the unpolarized cross section shown in Fig.~\ref{fig4}, the 
asymmetry is somewhat smaller at NLO than at LO, showing that
inclusion of NLO QCD corrections is rather important for the
analysis of the data in terms of $\Delta g$. 
%
%

We also conclude from the figure that there are excellent prospects for 
determining $\Delta g(x)$ from $A_{LL}^{\pi^0}$ measurements at
RHIC: the asymmetries found for the two different sets of polarized 
parton densities, which mainly differ in the gluon density, show
marked differences, much larger than the expected statistical
errors in the experiment, indicated in the figure. The latter
may be estimated by the formula
\begin{equation} \label{error}
\delta A_{LL}^{\pi} = \frac{1}{P^2\sqrt{{\cal L}\sigma_{\rm bin}}} \; ,
\end{equation}
where $P$ is the polarization of one beam, ${\cal L}$ the integrated
luminosity of the collisions, and $\sigma_{\rm bin}$ the unpolarized
cross section integrated over the $p_T$-bin for which the error is to be
determined. We have used the very moderate values $P=0.4$ and
${\cal L}=7$/pb, which are targets for the coming run. 
As mentioned above, we also take into account that at present
with the {\sc Phenix} experiment
a pion measurement is possible only over half the azimuthal angle.

\section{Conclusions}
%
We have presented in this paper the complete NLO QCD corrections for the 
partonic hard-scattering cross sections relevant for the
spin asymmetry $A_{LL}^{\pi}$ for high-$p_T$ pion production 
in hadron-hadron collisions. This asymmetry is a promising
tool to determine the spin-dependent gluon density in the nucleon
and will be measured in the coming run with polarized protons at RHIC. 
Our calculation is based on a largely analytical evaluation 
of the NLO partonic cross sections. 

We found that the NLO corrections to the polarized cross section
are somewhat smaller for RHIC than those in the unpolarized case. The
polarized cross section shows a significant reduction of scale
dependence when going from LO to NLO. Upcoming RHIC data should
be able to provide first information on $\Delta g$ even for rather
moderate integrated luminosities.

\section*{Note added:}
%
While nearing completion of our work, we learned that D.\ de Florian
has performed~\cite{ref:ddfnew} the same calculation, using the 
``Monte-Carlo'' method outlined in Sec.~2. This provides an extremely 
welcome opportunity for comparing the results. Early comparisons show 
very good agreement of the numerical results.

\section*{Acknowledgments}
%
We are grateful to D.\ de Florian for many valuable discussions
and for comparison with his results. B.J.\ and M.S.\ thank the RIKEN-BNL Research Center 
and Brookhaven National Laboratory for hospitality and support during the
final steps of this work.
B.J.\ is supported by the European Commission IHP program under contract number
HPRN-CT-2000-00130. 
W.V.\ is grateful to RIKEN, Brookhaven National Laboratory and the U.S.\
Department of Energy (contract number DE-AC02-98CH10886) for
providing the facilities essential for the completion of this work.

%

%
\end{document}